\documentclass[a4paper,aps,amsmath,amssymb,showpacs,10pt,twocolumn,floatfix,superscriptaddress]{revtex4}
\usepackage{graphicx}
\usepackage{graphics,color}
\usepackage{umlaut}

\newcommand{\bs}{\;\;\;\;\;}

\newcommand{\ve}{\mathbf}

\begin{document}

\title{Bound hole states in a ferromagnetic (Ga,Mn)As environment}
\author{M.J. Schmidt}
\affiliation{Physikal. Institut, Experimentelle Physik III, Universität Würzburg, Am Hubland, 97074 Würzburg, Germany}
\affiliation{Institut f. Theoretische Physik, Universität Würzburg, Am Hubland, 97074 Würzburg, Germany}

\author{K. Pappert}
\affiliation{Physikal. Institut, Experimentelle Physik III, Universität Würzburg, Am Hubland, 97074 Würzburg, Germany}

\author{C. Gould}
\affiliation{Physikal. Institut, Experimentelle Physik III, Universität Würzburg, Am Hubland, 97074 Würzburg, Germany}

\author{G. Schmidt}
\affiliation{Physikal. Institut, Experimentelle Physik III, Universität Würzburg, Am Hubland, 97074 Würzburg, Germany}

\author{R. Oppermann}
\affiliation{Institut f. Theoretische Physik, Universität Würzburg, Am Hubland, 97074 Würzburg, Germany}

\author{L.W. Molenkamp}
\affiliation{Physikal. Institut, Experimentelle Physik III, Universität Würzburg, Am Hubland, 97074 Würzburg, Germany}

\date{\today}

\begin{abstract}
A numerical technique is developed to solve the Luttinger-Kohn equation for impurity states directly in k-space and is applied to calculate bound hole wave functions in a ferromagnetic (Ga,Mn)As host. The rich properties of the band structure of an arbitrarily strained, ferromagnetic zincblende semiconductor yields various features which have direct impact on the detailed shape of a valence band hole bound to an active impurity. The role of strain is discussed on the basis of explicit calculations of bound hole states.

\end{abstract}

\maketitle

\section{Introduction}
Over the past decade, Mn-doped III-V semiconductors have been investigated by many groups. So far, some properties of those materials are not completely understood. At high doping concentrations ferromagnetic behavior is observed. The ferromagnetism can be well explained by a mean-field Zener model \cite{ohno,abolfath,dietl} which accounts for the exchange of the Mn d-electrons in combination with the usual III-V semiconductor band structure. On the other hand, in the very dilute limit, the material is non-magnetic, with single valence band holes bound to Mn-acceptors. Previously, this diluted doping regime has been modelled using $\ve k.\ve p$-theory based on the spherical model of Baldereschi and Lipari \cite{bhattacharjee,baldereschi} and in a tight-binding-approximation \cite{flatte}. However, the crossover between the opposite limits of high and low doping regimes has hardly been investigated theoretically. There are several situations where this crossover regime is important. For example, due to disorder, spatially separated low-hole-density islands with localized hole states may occur inside a ferromagnetic host material. Moreover in patterned structures, such as e.g. the tunneling structure in Ref. \cite{pappert}, one must be able to describe a system where both, localization and ferromagnetism, are important at the same time.

In this paper we investigate localized hole states in ferromagnetic $\rm Ga_{1-x}Mn_xAs$. The usual approach to describe single impurity states at the Mn-sites starts from the bulk band structrue of non-magnetic GaAs and incorporates the magnetic moment of an isolated Mn-impurity by a spin-dependent local potential \cite{flatte}. In the crossover regime, which we aim to describe here, the ferromagnetic order in the host material should also be accounted for. We incorporate this by describing the bulk $\rm Ga_{1-x}Mn_xAs$ host using the Zener model of \cite{dietl,abolfath}. Since the magnetic properties of the acceptor are already included in the bulk description, we can resort to a simple Coulomb potential without spin-dependence to describe the Mn-impurity.

We have developed a numerical technique to solve the Luttinger-Kohn equation \cite{luttkohn} for impurity states directly in $\ve k$-space and in the presence of an arbitrarily complex host band structure. Our technique it sufficiently fast to avoid the need for using the spherical approximation \cite{baldereschi}. Therefore all features of the $\ve k.\ve p$ band structure of $\rm Ga_{1-x}Mn_xAs$ (e.g. band warping) are taken into account. In this paper, a hydrogen-like impurity in strained ferromagnetic $\rm Ga_{1-x}Mn_xAs$ is considered, but the technique itself is restricted neither to a Coulomb potential nor to a particular ferromagnetic semiconductor.

\section{The model}
The general stationary Schrödinger equation for describing an impurity state $\left|\psi\right>$ inside a host material has the form
\begin{equation}
\left[ H_0(\hat{\ve k},\ve M,\epsilon) + U(\hat{\ve x})\right] \left|\psi\right> = E_b\left|\psi\right>\label{schroedingerequation}
\end{equation}
where $E_b$ is the binding energy of the hole bound by an attractive potential $U(\hat{\ve x})$ originating from the charge of the impurity. $H_0(\hat{\ve k},\ve M,\epsilon)$ describes the host material band structure. For our present purposes is suffices to use a six band $\ve k.\ve p$-model for $H_0$. Since it is diagonal in $\ve k$-space, $H_0$ can be written by the $\ve k$-dependent $6\times 6$-matrix
\begin{equation}
H_0(\ve k,\ve M,\epsilon)=H_{LK}(\ve k)+H_{so}+H_{pd}(\ve M)+H_s(\epsilon) \label{h0}.
\end{equation}

Below, we describe briefly each constituent of $H_0$; the interested reader can find their explicit forms in the literature \cite{abolfath,dietl,luttkohn}.

- The Luttinger-Kohn Hamiltonian \cite{luttkohn} $H_{LK}(\ve k)$ describes the kinetic energy of a hole in a crystal environment. It is proportional to $\frac{|\ve k|^2}{m_{\rm eff}(\ve k/|\ve k|)}$ (with $m_{\rm eff}$ the effective mass in a given direction) and thus yields purely parabolic bands. The effective mass in each direction (defined as the inverse band curvature at the $\Gamma$-point) is determined by the three experimentally obtained Luttinger-parameters $\gamma_1,\gamma_2,\gamma_3$.

- The spin-orbit Hamiltonian \cite{luttkohn} $H_{so}$ is constant in $\ve k$-space. It separates the bands with a total angular momentum $j=\frac12$ (split-off bands) from the bands with $j=\frac32$ by $\Delta_{SO}=0.34eV$. Because of the relatively large splitting, the $j=\frac12$ bands are sometimes neglected, resulting in a four-band model. We do not use this approximation and work with the full six-band model.

- The mean-field pd-exchange Hamiltonian \cite{abolfath,dietl} $H_{pd}(\ve M)$ is also constant in $\ve k$-space but depends on the magnetization orientation. It can be written as $H_{pd}(\ve M)=6B_g \hat{\ve M}\cdot \ve S$ where $\hat{\ve M}=\ve M/|\ve M|$ is the mean direction of all Mn d-spins and $\ve S$ is the valence band hole-spin operator. This Hamiltonian breaks time-reversal symmetry and thus lifts the two-fold Kramers-degeneracy. Together with the spin-orbit coupling it completely lifts the degeneracy at $\Gamma$. In this paper, we use $B_g=15meV$.

- The strain Hamiltonian \cite{birpikus} $H_s(\epsilon)$ depends on the strain tensor $\epsilon$. In this paper we only investigate non-sheared crystals ($\epsilon_{xy}=\epsilon_{yz}=\epsilon_{zx}=0$). If necessary, shear-strain can be treated on the same footing. Since high-quality $\rm Ga_{1-x}Mn_xAs$ is always grown pseudomorph on a substrate with a slightly different lattice constant, the most important strain is growth strain. Additional in-plane strains arising from partial relaxation \cite{stripes} and symmetry breaking effects \cite{pappert} can be observed and are investigated in this paper.

- As a basis we initially work with the total angular momentum eigenstates $\left|j,m_j\right>$ where $m_j$ is the magnetic quantum number with respect to a quantization axis in the [001]-direction. We refer to this basis in the following as the 'jm-basis'. 

\bigskip

\begin{figure}
\centering
\includegraphics[width=240pt]{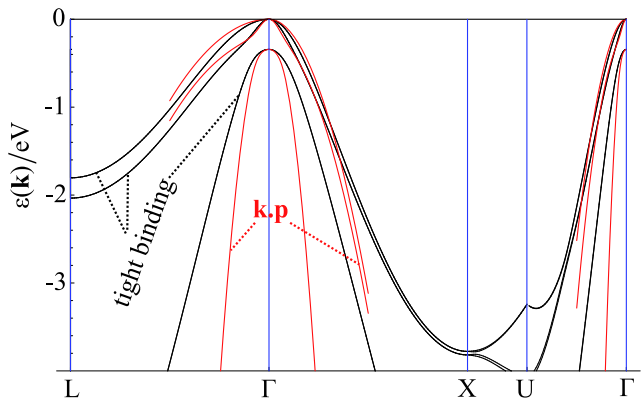}
\caption{Pure GaAs band structure of the $\Gamma_7$ and $\Gamma_8$ bands calculated with two different methods: The lines which are present in the whole Brillouin-zone (black-online) result from a tight binding calculation \cite{chadi} while the lines which are only plotted near $\Gamma$ (red-online) are obtained from a $\ve k.\ve p$-type calculation.}
\label{tightbindingkp}
\end{figure}

The main contribution to the wave function comes from the small $\ve k$ regime of the Brillouin-zone. For this reason, we use the $\ve k.\ve p$-approximation, which is exact to order $k^2$, and thus is best suited in this domain. This makes it a good approach to describe holes or electrons in semiconductors where all important electronic properties refer to the band edges and a high numerical accuracy is essential. To check the quality of the $\ve k.\ve p$-method deep inside the Brillouin zone we compare the band structures of pure GaAs calculated in $\ve k.\ve p$ and in tight binding \cite{chadi}. Figure \ref{tightbindingkp} shows that the top four bands nearly coincide. We will show later that the contribution of the split-off bands to an impurity state is very small so that we do not expect this difference in the two methods to lead to a significant effect on the results. 

\begin{figure}
\centering
\includegraphics[width=210pt]{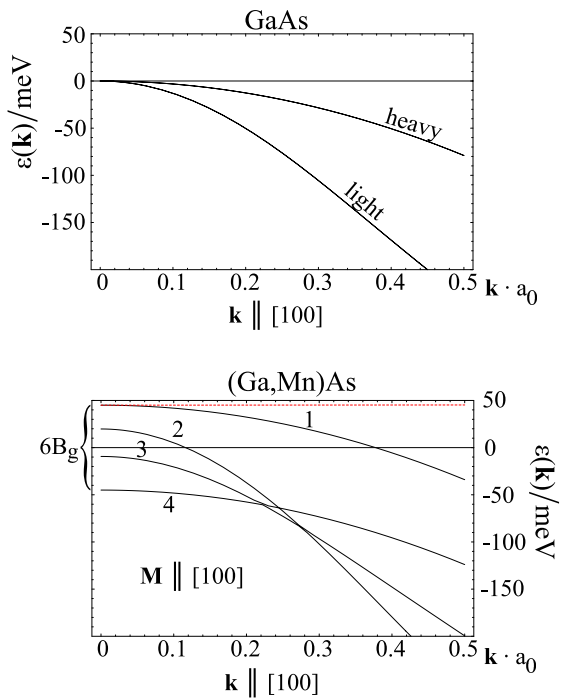}
\caption{The $j=3/2$ valence bands of ferro- and non-magnetic materials. The top figure shows the typical non-magnetic GaAs band structure with heavy and light holes. The bottom figure is the typical ferromagnetic band structure of (Ga,Mn)As. The bands are labeled by numbers. The total spin-splitting at $\Gamma$ is $6B_g$. The dotted line (red-online) is the energy level to which the binding energy of the bound hole is referred.}
\label{pdsplit}
\end{figure}

One central point of our argumentation is that the host crystal is a ferromagnet and therefore has a completely different band structure than a non-magnetic material. Since the wave function of a bound hole is strongly influenced by the underlying band structure, the knowledge of those differences is important to understand the results of our calculations. The ferromagnetism of the host material is modelled by the mean-field pd-exchange Hamiltonian $H_{pd}(\ve M)$, which couples directly to the hole spin. In figure \ref{pdsplit} we show the difference between a non-magnetic (e.g. GaAs) and a ferromagnetic (e.g. (Ga,Mn)As) band structure. While in GaAs the top subbands of the valence band (i.e. the bands with $j=3/2$) are degenerate at $\Gamma$, this degeneracy is completely lifted by $H_{pd}$. At $\Gamma$, $H_{pd}$ separates the pure spin-up and spin-down (quantized in $\ve M$-direction) bands by $6B_g$. This energy separation leads to extremely anisotropic effective masses for small $\ve k$ and the usual labeling of 'heavy' and 'light' hole bands fails due to the complicated band mixing caused by the mean-field pd-exchange. Therefore we simply assign a number to label each band.

\section{Self-consistency equation for impurity states}
The explicit $\ve k.\ve p$-form \cite{luttkohn} of the stationary Schrödinger equation (\ref{schroedingerequation}) for a valence band hole trapped by an impurity potential $U(\ve r)$ inside a host crystal is
\begin{equation}
\sum_{n'} H_{0}^{nn'}(\ve k)B_{n'}(\ve k)+\int d^3\ve k' \mathcal U(\ve k-\ve k')B_n(\ve k') = E_b B_n(\ve k)\label{origluttingerkohnequation}
\end{equation}
where $H_{0}(\ve k)$ is the $\ve k.\ve p$-approximated band structure Hamiltonian of $\rm Ga_{1-x}Mn_xAs$ \cite{dietl,abolfath}. $\mathcal U(\ve k)$ is the fourier transform of $U(\ve r)$ and $E_b$ is the binding energy of the impurity-hole system. $\ve B(\ve k)=(B_1(\ve k),...,B_6(\ve k))$ is the envelope wavefunction of the bound hole in $\ve k$-space. Once $\ve B$ is known, the actual wavefunction in real space can be obtained from the Fourier transform
\begin{equation}
\psi(\ve r)=\sum_n \int d^3\ve k \; B_n(\ve k) e^{i\ve k\cdot\ve r} u_{n}(\ve r)
\end{equation}
where $u_{n}(\ve r)=u_{j,m_j}(\ve r)$ is the lattice-periodic Bloch-function corresponding to the $\left|j,m_j\right>$ state. When $H_0$ describes only one simple s-like band with a single isotropic effective mass and $U(r)\propto\frac1r$ is a Coulomb potential, equation (\ref{origluttingerkohnequation}) reduces to the non-relativistic hydrogen problem transformed to $\ve k$-space. We have used this property to test the numerical procedure by application to this simpler problem which allows an analytical solution.

We are particularly interested in the ground-state wavefunction of a single valence band hole bound to a negatively charged Mn impurity, which we treat as a point charge in first approximation. Rather than using an unscreened Coulomb potential we start with a general Yukawa potential to eliminate the singularity of the Fourier transform of $\frac1r$ at $\ve k=0$. In case of the Yukawa interaction, the singularity is cut off by a screening parameter $\lambda$. Up to a coupling-constant we have
\begin{equation}
U(r)=\frac{e^{-\lambda r}}{r},\bs \mathcal U(k)=\frac1{2\pi^2}\frac1{k^2+\lambda^2}
\end{equation}
as the impurity potential. The limit $\lambda\rightarrow0$ yields the Coulomb interaction. We will investigate the Coulomb limit numerically by checking the convergence with respect to small but finite $\lambda$.

In order to obtain a numerical solution of equation (\ref{origluttingerkohnequation}) we rewrite it in a self-consistent form
\begin{widetext}
\begin{equation}
B_n(\ve k)=\frac1{E_b-H_{0}^{nn}(\ve k)}\left[\sum_{n'\neq n}H_{0}^{nn'}(\ve k)B_{n'}(\ve k)+\int d^3\ve k' \mathcal U(\ve k-\ve k')B_n(\ve k') \right]\label{recursionrelation}
\end{equation}
\end{widetext}
which can be solved by iteration. After each step, the binding energy $E_b$ is adjusted in order to obtain a self-consistent pair $(E_b,\ve B(\ve k))$ which solves the Schrödinger-equation (\ref{origluttingerkohnequation}).

Note that the $\int d^3\ve k' \mathcal U(\ve k-\ve k')B(\ve k')$ term requires a sum over all $\ve k'$ points for each $\ve k$ point. This means that for $N$ sampling points in $\ve k$-space each self-consistency step requires $\mathcal O(N^2)$ operations. We have designed a dedicated code which analytically integrates out parts of the angular integral \cite{angularintegral} and reduces the remaining integral to a matrix-multiplication in order to cope with this task.

The method described above can in principle be applied to any system with valence band holes bound to impurities in a ferromagnetic host, by setting the proper material parameters. In the remainder of this paper, we investigate the specific material system $\mathrm{ Ga_{1-x}Mn_xAs}$ in its ferromagnetic regime ($x\simeq 0.05$).

Valence band holes bound to Mn-impurities have been investigated in several papers \cite{bhattacharjee,flatte,baldereschi,linnarsson}. In all of these papers, the Mn-impurity has been treated as an isolated impurity center in a non-magnetic host. Here, we investigate the Mn-center inside a {\it ferromagnetic host}, using the mean-field pd-exchange Hamiltonian to describe the ferromagnetism of the host material. Since the fourfold degeneracy of the zinc-blende valence band at the $\Gamma$-point is lifted by the ferromagnetism, the band structure and the bound hole wave function can no longer be approximated by the spherical model of \cite{baldereschi}.

This deviation of the present model from the traditional treatment of impurity states complicates a direct comparison of the results of the calculations. In the limit of vanishing ferromagnetism ($B_g\rightarrow0$), however, the binding energy of Baldereschi and Lipari \cite{baldereschi} is recovered, which is a check of the validity of the numerical algorithm. For $B_g\neq 0$ we measure the binding energy relative to the top of the split valence band at $\Gamma$ (see fig. \ref{pdsplit}). Note however that this energy level is $3B_g$ higher than the GaAs band edge due to a total splitting between $\uparrow$- and $\downarrow$-spin of $6B_g\simeq90meV$. This fact should always be kept in mind when comparing the present results to non-magnetic impurity binding energies.

Because the $B_g\rightarrow0$ limit of the binding energy equals the Baldereschi-Lipari (BL) binding energy \cite{baldereschi}, the same criticism about the binding energy (too small as compared with experiment) as for the BL-model is applicable here. To obtain binding energies that can be directly compared with experiments, we can follow the same path as BL, i.e. including a central cell correction. We have not done this yet, but we expect the central cell correction to mainly manifest itself in an overall scale factor of the wave function extent. The main results of this paper (i.e. the shape of the wave function) should remain unchanged by the central cell correction.

\section{Bound hole wave functions}
\begin{figure}
\centering
\includegraphics[width=250pt]{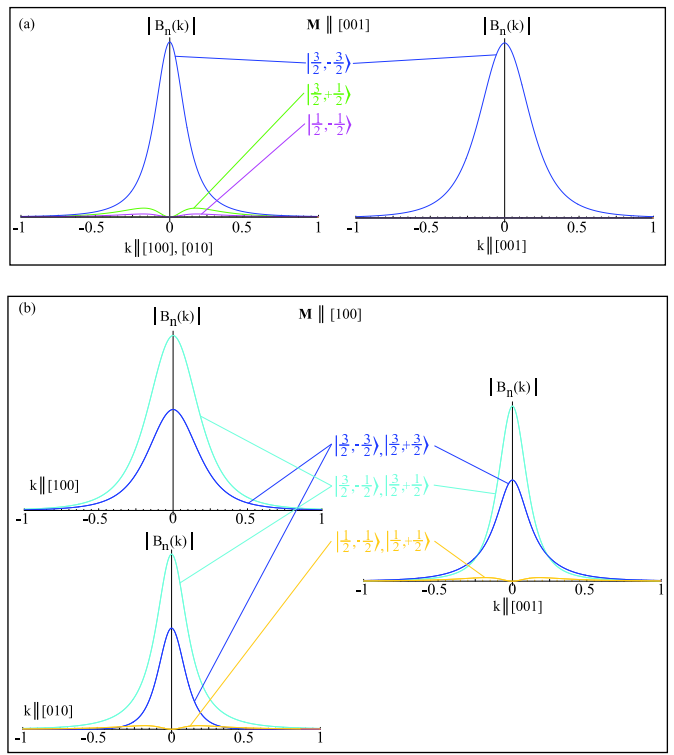}
\caption{Absolute values of envelope wavefunctions in $\ve k$-space in jm-representation for $\ve k \parallel [100],[010],[001]$ with (a) magnetization along [001] and (b) magnetization along [100]. The results are obtained from a calculation with 50 $\ve k$-space directions and 200 points in each direction. The abscissa ($k$-axis) is given in units of $a_0^{-1}$ with $a_0=0.565\;nm$ the lattice constant of GaAs.}
\label{wavefunctions}
\end{figure}

The numerical solution of Eq. (\ref{recursionrelation}) results in an envelope wave function of a bound hole, given as a 6-component vector $\ve B(\ve k)$ in $\ve k$-space. The basis in which $\ve B$ is represented can be chosen by convenience. During the self-consistency cycle, the usual jm-basis is used. For analyzing the result, however, a different basis is more convenient. In figure \ref{wavefunctions} we show the $\ve k$-space impurity wave function inside un-strained (Ga,Mn)As in the usual jm-representation. In part (a) the magnetization is oriented along [001] and in part (b) along [100]. Since these two magnetization directions are equivalent, it should be possible to map both results (a) and (b) onto each other. The use of jm-representation, however, implies the choice of a $j$-quantization axis. This choice prevents a trivial mapping and makes the jm-basis unattractive for visualization purposes.

\begin{figure}
\centering
\includegraphics[width=250pt]{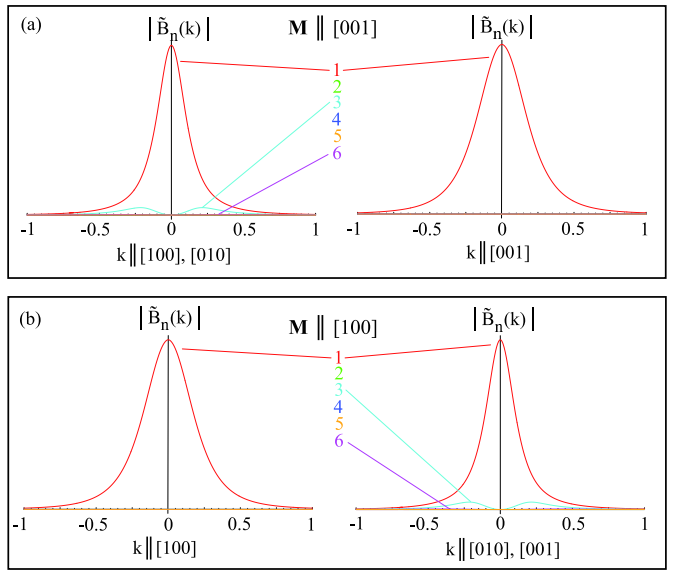}
\caption{Absolute values of envelope wavefunctions in $\ve k$-space in eigenbasis representation. As in figure \ref{wavefunctions} the magnetization orientation is [001] in (a) and [100] in (b). Here the different lines represent different bands. These bands are enumerated as in figure \ref{pdsplit}}
\label{eigenwavefunctions}
\end{figure}

In order to obtain a comparison that is independent of the quantization axis, we represent $\ve B$ in the $\ve k$-dependent basis in which the band structure Hamiltonian $H_0(\ve k)$ is diagonal \cite{numericalbasis}. Thus, a $\ve k$-dependent basis transform $T(\ve k)$ is performed, so that
\begin{equation}
T(\ve k) H_0(\ve k) T^{\dagger}(\ve k)=\left(\begin{matrix}
\epsilon_1(\ve k) & & 0\\
& \ddots & \\
0 & & \epsilon_6(\ve k) 
                                             \end{matrix}\right)
\end{equation}
where $\epsilon_i(\ve k)$ is the dispersion of the $i$th band. The band order is chosen such that $\epsilon_i(0)>\epsilon_j(0)$ for $i<j$. The transformed wave function $\tilde{\ve B}(\ve k)=T(\ve k)\ve B(\ve k)$ is from now on referred to as the eigenbasis-representation of the envelope wave function. Visualized in the eigenbasis-representation (see fig. \ref{eigenwavefunctions}), the results can easily be seen to obey the proper symmetry with respect to magnetization orientation.

Another important reason for using the eigenbasis-representation of the wave function is that it makes the connection between the wave function and the host band structure explicit: The $i$th component of $\tilde{\ve B}(\ve k)$ refers directly to the $i$th band of the host band structure at $\ve k$. This means that by representing $\ve B$ in eigenbasis one can directly see from which bands the wave function derives.

From figure \ref{eigenwavefunctions} it is evident that the main contribution of the bound hole wave function derives from the first band (see fig. \ref{pdsplit}), which is energetically highest at the $\Gamma$-point. This is a very general feature of the present model and results from the large spin-splitting caused by the pd-exchange Hamiltonian $H_{pd}$. This property of the impurity state is a direct implication of the treatment of the host material as a ferromagnet. It also implies that the spin of the hole is also mainly determined by band 1, i.e. it is antiparallel to the spin of the Mn d-electrons. Because of band mixing, this statment cannot be exact; from our calculations, we estimate the bound hole spin $s_z=-0.472$ for magnetization along $z$-direction \cite{antispin}.

\begin{figure}
\centering
\includegraphics[width=230pt]{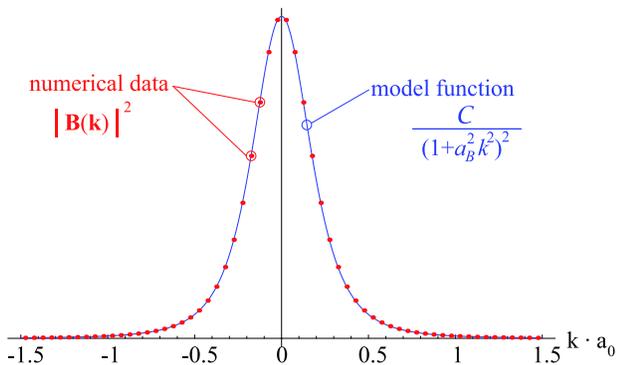}
\caption{Fit of a hydrogen ground state wave function model (blue online) to the numerical data of $|\ve B(\ve k)|^2$ (red online) for $\ve k$ in magnetization direction.}
\label{bohrFit}
\end{figure}

The 3D fourier transform of a spherically symmetric function is mathematically trivial. The transformation of our $\ve k$-space wave function $\ve B$ to real space, however, is more challenging, due to the broken spherical symmetry. In order to obtain the shape of the bound hole wavefunction in real space we make use of the similarity to the hydrogen problem: the ground-state Coulomb wave function can be transformed to $\ve k$-space easily. It is spherically symmetric and its radial part is
\begin{equation}
B^{(h)}(k)=\frac{\mathcal C}{(1+a_B^2k^2)^2}
\end{equation}
where $a_B$ is the Bohr radius and $\mathcal C$ a normalization constant. To extract the extent of the wave function in a given direction in real space, $B^{(h)}(k)$ can be fitted directly to the numerical data of $\ve B(\ve k)$ for $\ve k$ in this direction. To be precise, we fit $\left(\sum_{n=1}^6 \left|B_n(\ve k)\right|^2\right)^{1/2}$ to the hydrogen wavefunction. Figure \ref{bohrFit} shows the high quality of such a fit, which is due to the use of a Coulomb potential and the similarity to the hydrogen problem. The Bohr radii can be used as an interpolation dataset to obtain the Bohr radius $a_B(\hat{\ve r})$ in an arbitrary real-space direction $\hat{\ve r}$. We visualize the wavefunction shape by plotting the 3D surface $|\ve r| = a_B(\hat{\ve r})$.

\begin{figure}
\centering
\hspace*{-0.4cm} \includegraphics[width=250pt]{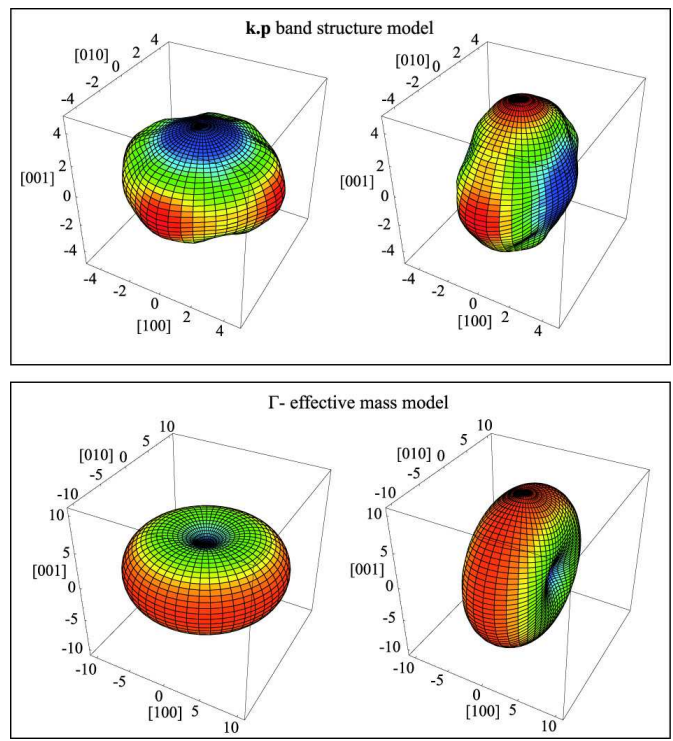}
\caption{Shape of bound hole wavefunctions. The top part is the result of a full $\ve k.\ve p$ band structure model and the bottom part results from a simple effective mass treatment. On the left side the magnetization is oriented along [001], on the right side the magnetization is oriented along [100]. The colors are chosen so that the smallest extent is blue and the largest extent is red. The axes are given in units of the GaAs lattice constant $a_0$.}
\label{shapeExample}
\end{figure}

\begin{figure}
\centering
\includegraphics[width=230pt]{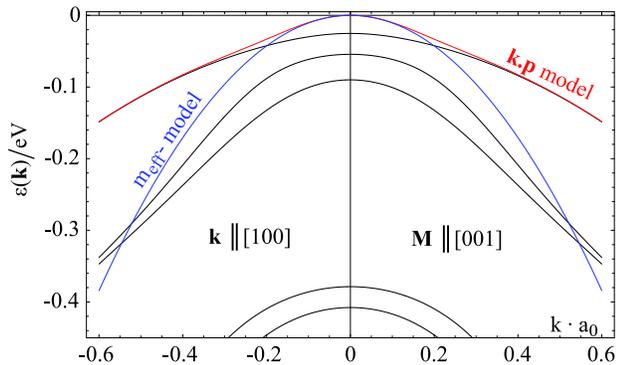}
\caption{(Ga,Mn)As band structure for $\ve k$ in [100]-direction and $\ve M$ in [001]-direction. The line labeled by $m_{\rm eff}$-model (blue-online) refers to the parabolic top band with curvature determined by the effective mass at the $\Gamma$-point and the line labeled by $\ve k.\ve p$-model (red-online) refers to the top band in full $\ve k.\ve p$-treatment. Note that the zero point of energy is chosen as the reference energy level of the binding energy (see also fig. \ref{pdsplit} and text near by).}
\label{bandsatgamma}
\end{figure}

Figure \ref{shapeExample} shows the resulting shape of the wave function in real space corresponding to figures \ref{wavefunctions} and \ref{eigenwavefunctions}. Obviously, the extent of the wave functions is largest perpendicular to the magnetization direction. The physical explanation for this can be found by an analysis of the host band structure, especially the direction-dependent effective masses at $\Gamma$: The Bohr radius of a hydrogen wave function of a particle with mass $m_{\rm eff}$ bound in a Coulomb potential which is screened by a dielectric constant $\kappa$ is given by
\begin{equation}
a_B=\frac{4\pi \kappa\epsilon_0 \hbar^2}{m_{\rm eff} e^2}.
\end{equation}
Setting $m_{\rm eff}$ to the effective mass of the top band in a particular direction and $\kappa\simeq12.4$ to the dielectric constant of GaAs gives a rough approximation for the shape of the wave function. The result of this approximation is shown in the lower row of figure \ref{shapeExample}. Obviously the oblate shape of the wave function following the magnetization orientation is a direct consequence of the ferro-magnetic band structure of the host material. In this simple model, only the $\Gamma$-point of the band structure is contributing to the hole wave function. The full model shows how details of the band structure away from $\Gamma$ modify the result. In particular, the extent perpendicular to the magnetization shrinks when taking into account the band structure away from $\Gamma$. This can easily be understood by considering the difference between a parabolic band with curvature determined by the effective mass at $\Gamma$ and the same band calculated in full $\ve k.\ve p$, as shown by figure \ref{bandsatgamma}: The approximation underestimates the effective mass in [100]-direction. Since $a_B \propto m_{\rm eff}^{-1}$, this underestimation leads to an overestimation of the wave function extent in this direction, as seen in figure \ref{shapeExample}.

The wave functions deviate strongly from the spherically symmetric result obtained using $\ve k.\ve p$-modelling by Bhattacharjee et al. \cite{baldereschi,bhattacharjee} for zero magnetization. We also obtain a much more pronounced anisotropy in the wave function than observed in the tight-binding model of \cite{jmtang}. These differences are direct consequences of the ferromagnetism of the host material. The fact that the wave-function shape follows the magnetization orientation is due to the pd-exchange \cite{jmtang} and the effect is amplified by the ferromagnetic band structure.

\section{Strain and Symmetry-breaking effects}
\subsection{Growth strain}
\begin{figure}[here]
\centering
\includegraphics[width=235pt]{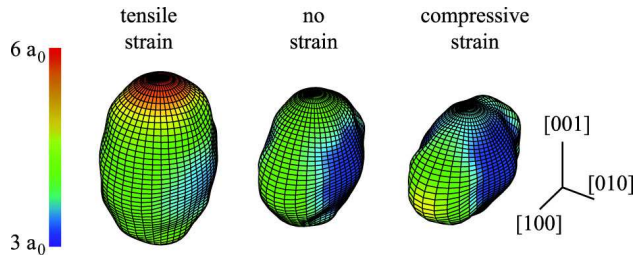}
\caption{The effect of tensile (left) and compressive (right) growth strain on the impurity wave functions. The non-strained wave function in the middle is for comparison purposes. The magnetization is along [010].}
\label{strains}
\end{figure}

(Ga,Mn)As grown epitaxially on GaAs is compressively strained in the growth-plane, leading to a reduction of the cubic symmetry and thus a change of magnetic easy axes. Calculations \cite{dietl,abolfath} and experiments have shown that unstrained (Ga,Mn)As has three magnetic easy axes, i.e. $\left<100\right>$, for hole densities typical for transport samples. A growth strain in one of these directions \cite{gould-easyaxes} makes this direction magnetically hard \cite{growthstrainnote}. Because of this, we restrict the following discussion to magnetization orientations along one of the easy axes [100] or [010], or in between, and choose [001] as the growth direction. In figure \ref{strains} we show the wave functions for tensile strain \cite{tensilefootnote}, for a non-strained and a compressively strained host.

Obviously, a compressive strain enhances the in-plane extent of the wave function in the magnetization direction while a tensile strain produces a larger extent in the growth direction. Again, this effect can easily be understood by investigating the dependence of the band structure on the strain properties of the host: A large tensile growth strain pushes the second and third bands to lower energies. This leads to a larger area of the Brillouin zone where the first band is not disturbed by the others. In this limit, the first band has light-hole character (large curvature at $\Gamma$) in the growth direction and heavy-hole character for in-plane directions. Since $a_B\propto m_{\rm eff}^{-1}$, this consideration qualitatively leads to the left wave function in figure \ref{strains}. The opposite limit of large compressive growth strain is slightly more complex. Here the second band becomes energetically close to the first band and strongly affects it. However, the effect of the second band is not the same for all $\ve k$-directions. In growth direction, it pushes the first band up so that it gets heavier and the wave function shrinks in growth direction. In the plane, instead, the first band gets nearer to the second and so the curvature increases, leading to an enhancement of the wave function extent in [100] direction.

\subsection{In-plane strain}
The treatment of growth-strain in (Ga,Mn)As is well known and straightforward \cite{dietl,abolfath,birpikus}. It breaks the symmetry between the [001] growth direction and each of the two in-plane easy axes ([100] and [010]), while the symmetry between [100] and [010] is retained. Experimentally \cite{tamr-letter,fingerprint}, however, symmetry is also broken between those two in-plane directions. No first principles explanation of this symmetry breaking has yet been established. Rather, the effect has been modeled successfully by a phenomenological in-plane strain term: Gould et al. \cite{tamr-letter} introduced a strain term of order 0.1\% and observed a magnetization orientation dependence of the partial DOS, leading to a remanent tunneling anisotropic magnetoresistance (TAMR) effect. Later, a very large amplification of this effect at low temperatures was observed \cite{vltamr-letter} and explained in terms of a magnetization orientation driven metal insulator transition (MIT) \cite{pappert}. Even without any strain, the shape of a bound hole wave function depends on the magnetization orientation.

A difference between the hole wave functions for magnetization along two equivalent directions (in this case [100] and [010] \cite{equivalentdirections}), however, can only be obtained by the introduction of a uniaxial symmetry breaking strain term along one of these directions - in analogy to what was needed to explain the remanent TAMR effect of reference \cite{tamr-letter}. This is one reason for investigating the effects such kinds of strain theoretically. Another reason is for example the strongly anisotropic in-plane strain resulting from partial relaxation in patterned (Ga,Mn)As samples \cite{stripes,huempfner,wunderlich}

A key ingredient of our investigation is therefore studying the effects of in-plane strain, which we refer to as $\epsilon_i\equiv \epsilon_{xx}$ in addition to the growth strain $\epsilon_g\equiv\epsilon_{zz}$; in the following we always use $\epsilon_g=0.005$, which is a reasonable value for (Ga,Mn)As layers with thicknesses in the 10nm range \cite{dietl,abolfath}. We want to argue that for such parameters a switch of the magnetization orientation between the two in-plane easy axes can trigger a metal insulator transition (MIT) in a region of low hole density. Consider (Ga,Mn)As on the insulating side of the MIT. The mean overlap between neighbouring holes bound to impurity centers is too small to allow the holes to occupy spatially extended states. Then, assume some mechanism which causes the bound hole wave functions to grow in size such that there is sufficient mean overlap to destroy the localization. This implies that the material has crossed the MIT and now has a metallic character. We show that one such mechanism is a magnetization reorientation from one in-plane easy axis to the other.

In order to support this claim, we investigate the total extent of the wave function in terms of the volume $V$ enclosed by the surface $|\ve r|= a_B(\hat{\ve r})$ and the extent of the wave function in the growth direction in terms of $a_B([001])$. The latter quantity is useful for investigating effects of tunneling in the growth direction ([001]) such as those in \cite{pappert}.

\begin{figure}
\centering
\includegraphics[width=230pt]{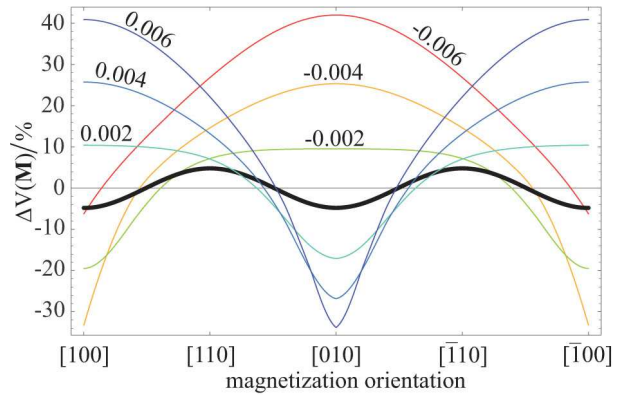}
\caption{Relative extent of the bound hole wave functions in dependence of magnetization orientation. Each line is labeled by a particular $\epsilon_i$. The thick black line is for $\epsilon_i=0$. $\Delta V$ is given relative to the mean volume of the $\epsilon_i=0$ wave function.}
\label{volumes}
\end{figure}

Figure \ref{volumes} shows the magnetization direction dependence of the wave function extent, measured in terms of $V$. For $\epsilon_i=0$ we find the usual 4-fold symmetry as expected. For non-vanishing $\epsilon_i$, a strong uniaxial contribution appears which makes the total extent for magnetization parallel to $\epsilon_i$ very different from the total extent for magnetization perpendicular to $\epsilon_i$. For $|\epsilon_i| \simeq |\epsilon_g|$, the ratio $r(\epsilon_g)=V([100])/V([010])$ is up to 2. A doubling in size of the bound hole states should easily be able to drive a material which is close to the MIT through this transition.

\subsection{Critical strain}
The shape of a bound hole wave function shows another interesting feature which only appears at - from a bulk point of view - exotic strain and pd-exchange parameters. At interfaces, however, where the exchange coupling may be reduced, or in patterned samples, where strong in-plane strains are present, these parameters are reasonable. To be specific, this feature appears whenever the two top bands of the ferromagnetic host crystal are degenerate at $\ve k=0$. This happens, for example, if we apply a negative in-plane strain $\epsilon_i\simeq -0.005$ i.e. equal in magnitude to the growth strain. We then observe a discontinous change of the shape and character of the wave function as a function of $\epsilon_i$. 

\begin{figure}
\centering
\includegraphics[width=250pt]{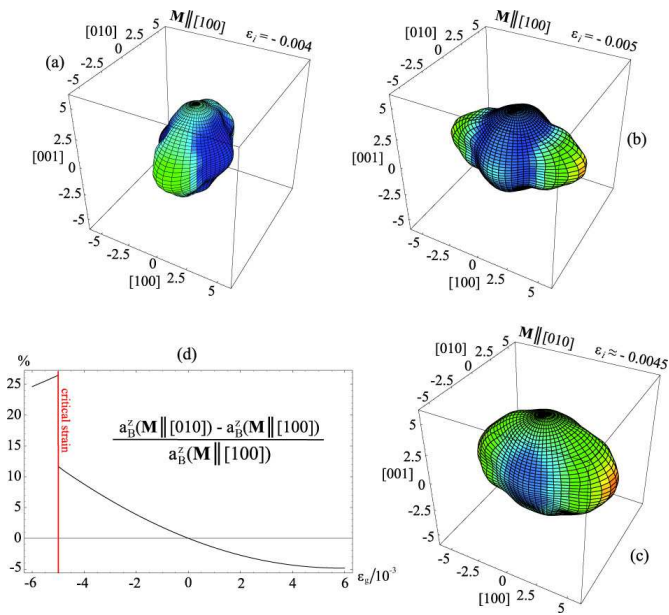}
\caption{Bound hole wave functions near critical strain: (a) and (b) show the wave functions for $\epsilon_i$ smaller than critical and larger than critical respectively with magnetization orientation along [100]. The wave function in (c) is for magnetization along [010] and strain near its critical value. (d) shows the relative difference of the extent of the bound hole wave function in growth direction ([001]) for magnetization in [100] and [010] direction.}
\label{critical}
\end{figure}

The discontinuity can best be seen from a comparison of the extent of the wave function in the growth direction $a_B^z$. In figure \ref{critical} (d) we show the $\epsilon_i$-dependence of the quantity 
$$\frac{a_B^z(\ve M\parallel [010])- a_B^z(\ve M\parallel [100])} {a_B^z(\ve M\parallel [100])}$$
where $a_B^z(\ve M\parallel [100])$ is the extent of the wave function in growth direction when the magnetization is along [100]. At small $\epsilon_i$ we find the behaviour which we would expect from figure \ref{volumes}, i.e. the extent is larger (smaller) for $\ve M$ along [100] then for $\ve M$ along [010] for positive (negative) $\epsilon_i$. At $\epsilon_i\simeq -0.005$, however, a sharp jump as a function of $\epsilon_i$ appears. We call the abscissa at which this jump appears the critical strain. Its exact value depends on other parameters like the pd-exchange coupling constant $B_g$ or the growth strain.

Not only the growth direction extent of a hole wave function is affected by this effects. The overall shape of the wave function changes discontinously as a function of $\epsilon_i$ at the critical strain, provided $\ve M\parallel [100]$. Figure \ref{critical} shows wave functions for $\epsilon_i$ near its critical value. (a) and (b) show the difference for both sides of the criticality. The extent in magnetization direction is large for $\epsilon_i$ above criticality, while the extent in the plane perpendicular to the magnetization is smaller. For magnetization along [010], however, there is no such sharp switch - (c) shows the shape of the wave function in this case. This means that for $\epsilon_i$ larger than its critical value the bound hole wave function no longer has an oblate shape following the magnetization orientation.

\begin{figure}
\centering
\includegraphics[width=250pt]{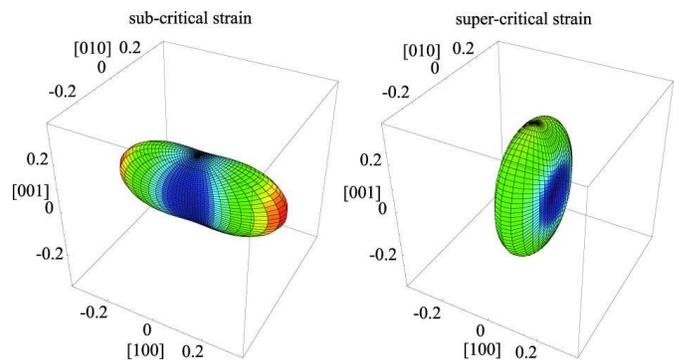}
\caption{$\ve k$-direction dependence of the effective mass of the first band at the $\Gamma$-point for magnetization along [100] on the two sides of the strain criticality. The axes are given in units of the free electron mass.}
\label{ems}
\end{figure}

Let us discuss the underlying principle of the concept of critical strain: As mentioned above, the bound hole wave function derives mainly from the first band, i.e. the band which is energetically highest at $\Gamma$. At critical strain and $\ve M\parallel [100]$, however, the first and second band are degenerate. In this situation, a small variation of $\epsilon_i$ lifts the degeneracy and the wave function 'switches' to one of the two bands - which one depends on the sign of the variation. The $\ve k$-dependence of these two bands is in general very different. In figure \ref{ems}, the most important charateristic for the bound hole wave function, i.e. the effective mass of the first band, is shown. Obviously, the shapes of the wave functions in figure \ref{critical} (a) and (b) are again approximately related to the sub- and super-critical strain effective masses by the relation $a_B\propto m_{\rm eff}^{-1}$. 

For magnetization along [010] there is no degeneracy at $\ve k=0$ of the first two bands at the critical strain. A small change in $\epsilon_i$ is therefore not able to change the band ordering at $\Gamma$ and no sharp switching occurs.

\section{Conclusion and outlook}
We have investigated valence band holes bound to active Mn-impurities inside (Ga,Mn)As in the ferromagnetic regime. Our main focus was the spatial extent of the bound hole in dependence of magnetization direction and strain properties of the host material. A numerical technique has been developed to solve the Luttinger-Kohn equation directly in $\ve k$-space for an arbitrarily complex host band structure. This technique is therefore not limited to (Ga,Mn)As but can be applied to many ferromagnetic systems which are describable within the $\ve k.\ve p$-approximation. The resulting wave functions differ strongly from the spherically symmetric results obtained by \cite{baldereschi,bhattacharjee}.

One of the main results is that the wave function derives mainly from the band of the host band structure which has highest energy at the $\Gamma$-point. The qualitative appearance of the bound hole wave function is an oblate shape which follows the magnetization orientation. The band ordering, which can be affected by external parameters like strain, is extremely important for the shape of the impurity state wave function. Changing the order of the bands at $\Gamma$, e.g. by strain, leads to large differences in the shape and in the mean extent of the wave functions. With the help of this mechanism a magnetization switched metal-insulator transition has been explained \cite{pappert}.

It would be interesting in the future to study how a central cell correction affects the wave function of the impurity states and their binding energy. Since this concept is normally incorporated by additional Gaussian corrections to the Coulomb-potential, the partial integrals entering our numerics are still feasible.

Experimentally, we have seen evidence for hopping-like transport in various ferromagnetic semiconductor devices. With the help of the detailed information about the wave function extent in arbitrary directions provided by the present model, we are now in a position to develop more detailed percolation models aimed at understanding the magnetoresistance properties of these devices. We anticipate that the present theory should allow to design novel (Ga,Mn)As devices with tailored magnetoresistance functionality.

The authors thank G.E.W. Bauer, T. Dietl and M. Flatt\'{e} for useful discussions. We acknowledge financial support from the EU (FP6-IST-015728, NANOSPIN) and DFG (SFB 410).


\begin{thebibliography}{99}
\bibitem{angularintegral} This is important because of the singular nature of the impurity potential.
\bibitem{numericalbasis} Of course, this basis is only known numerically.
\bibitem{growthstrainnote} By growth strain in a particular direction we mean, that this direction is perpendicular to the (Ga,Mn)As layer. This biaxial compressive strain can be transformed to a uniaxial tensile strain which is perpendicular to the growth plane.
\bibitem{tensilefootnote} The easy axes in tensile strained (Ga,Mn)As may be oriented along the growth direction but we choose an in-plane magnetization anyway.
\bibitem{equivalentdirections} By equivalent, we mean that the directions can be mapped by operations from the corresponding symmetry group.
\end{thebibliography}
\end{document}